\documentclass[a4paper,11pt]{article} 
\usepackage{jheppub}
\usepackage{graphicx}
\usepackage{epsfig}
\usepackage{color} 
\usepackage{amssymb}
\usepackage{amsmath}
\usepackage{doi}
\usepackage{dsfont}
\usepackage{subcaption}
\usepackage{float}
\usepackage{mathtools}
\usepackage{verbatim}
\usepackage{slashed}
\usepackage[normalem]{ulem} 
\usepackage{xspace}

\usepackage[dvipsnames]{xcolor}

\newcommand{\mr}{\mathrm}

\newcommand{\hjjj}{H+3~\text{jets}}

\newcommand{\ga}{\gamma}
\newcommand{\hajj}{H\gamma+2~\text{jets}}

\newcommand{\muf}{\mu_\mr{F}}
\newcommand{\mur}{\mu_\mr{R}}
\newcommand{\xif}{\xi_\mr{F}}
\newcommand{\xir}{\xi_\mr{R}}

\newcommand{\VBFNLO}{{\tt{VBFNLO}}}
\newcommand{\SHERPA}{{\tt{SHERPA}}}

\newcommand{\MG}{{\tt{MadGraph}}} 
\newcommand{\MGNLO}{{\tt{MadGraph5\_aMC@NLO}}} 
 
\newcommand{\POWHEG}{{\tt{POWHEG}}}
\newcommand{\PBOX}{{\tt{POWHEG~BOX}}}

\newcommand{\PBOXRES}{{\tt{POWHEG~BOX~RES}}}

\newcommand{\PYTHIA}{{\tt{PYTHIA}}}

\newcommand{\PYTHIAE}{{\tt{PYTHIA8}}}
\newcommand{\HERWIG}{{\tt{HERWIG}}}
\newcommand{\HERWIGS}{{\tt{HERWIG7}}}
\newcommand{\VINCIA}{{\tt{Vincia}}}

\newcommand{\beq}{\begin{equation}}
\newcommand{\eeq}{\end{equation}}
\newcommand{\bea}{\begin{eqnarray}}
\newcommand{\eea}{\end{eqnarray}}

\newcommand{\gev}{\mr{GeV}}
\newcommand{\mev}{\mr{MeV}}

\newcommand{\bb}{b\bar b}

\newcommand{\ed}{\end{document}}

\title{NLO QCD and parton-shower effects for Higgs-boson production in association with a hard photon via vector-boson fusion }
\preprint{}

\newcommand{\TUBaff}{Institute for Theoretical Physics, University of T\"ubingen, Auf der Morgenstelle 14,
72076 T\"ubingen, Germany}

\author[]{Barbara J\"ager,}%
\author[]{Simon Reinhardt}%

\affiliation[]{\TUBaff}

\abstract{ We present an implementation of Higgs-boson production in association with a hard, isolated photon via vector-boson fusion in the framework of the \PBOX{} for the consistent matching of next-to-leading order QCD corrections with parton showers.
The impact of parton-shower settings and non-perturbative effects on Higgs observables is studied and found to be small, while larger corrections are found for distributions of the sub-leading jets. 
Various approaches for the isolation of the photon are explored. For typical setups, the isolation strategy is found to have little impact on even the most sensitive observables.  
  }


\begin{document}

\maketitle

\section{Introduction}
After the discovery of a particle compatible with the Higgs boson of the Standard Model (SM) at the CERN Large Hadron Collider (LHC), a precise determination of its properties is of prime relevance. An extraction of its coupling to bottom quarks is particularly challenging in the dominant Higgs production modes, such as gluon fusion and vector-boson fusion (VBF), because of overwhelming QCD backgrounds. However, as pointed out in~\cite{Gabrielli:2007wf}, the dominant background to the VBF production mode in the $H\to \bb$ decay mode, consisting of two light jets and a $b\bar b$ pair, can be efficiently suppressed by the requirement of an additional hard, central photon while reducing the VBF-induced signal cross section to a much lesser extent. Therefore, the signal-to-background ratio of the VBF process in the $H\to\bb$ decay mode is drastically improved. 
Moreover, the requirement of an extra photon in the central-rapidity region enhances the relative relevance of charged-current contributions and thus the sensitivity to the $WWH$ coupling relative to the $ZZH$ coupling probed in neutral current contributions to the VBF cross section.  
The electroweak (EW) process of Higgs boson production in association with a hard, central photon and two forward tagging jets, i.e.\ the process $pp\to\hajj$ which at tree-level involves no strong coupling,  is thus expected to provide valuable information on properties of the Higgs boson that are difficult to access in other channels. 
More recently, this process has also been considered in the context of the Standard Model Effective Field Theory (SMEFT) as a means to constrain SMEFT operators~\cite{Biekotter:2020flu}. 

An indispensable prerequisite for a precise determination of Higgs couplings in this channel at the LHC are state-of-the art tools for the simulation of the VBF-induced $\hajj$ process. A parton-level Monte-Carlo program including the next-to-leading order (NLO) QCD corrections to this reaction has been developed in~\cite{Arnold:2010dx} and implemented in the publicly available \VBFNLO{} program~\cite{Arnold:2008rz}. Experimental analyses performed by the ATLAS collaboration with 30.6~fb$^{-1}$~\cite{ATLAS:2018jvf} and 132~fb$^{-1}$\cite{ATLAS:2020cvh} of data collected at a center-of mass (c.m.s) energy of $\sqrt{s}=13$~TeV have been based on simulations of the signal process with the \MGNLO{} generator~\cite{Alwall:2014hca} run at leading order (LO) and combined with \PYTHIAE{} in the earlier work and at NLO with \HERWIGS{} in the later one.  

However, a dedicated implementation in the framework of the \PBOX{}~\cite{Alioli:2010xd} which allows for a matching according to the \POWHEG{} prescription~\cite{Nason:2004rx,Frixione:2007vw} with \PYTHIA{}~\cite{Sjostrand:2006za,Sjostrand:2007gs} as well as \HERWIG~\cite{Corcella:2000bw,Bellm:2015jjp}, and thus facilitating a systematic estimation of ambiguities due to the parton-shower generator beyond the intrinsic uncertainties of each program, has not existed until now. With the work presented in this article we wish to close that gap. 
    

The paper is structured as follows: We provide some details on the definition of the VBF-induced $\hajj$ process and its implementation in the \PBOX{} in Sec.~\ref{sec:implementation}. In Sec.~\ref{sec:pheno} we apply the newly developed Monte-Carlo generator to phenomenological studies with particular emphasis on strategies for an isolation of the direct photon production component. Our conclusions are presented in Sec.~\ref{sec:conclusions}.

\section{Details of the implementation}
\label{sec:implementation}
The production of a $\hajj$ final state in hadronic collisions can, in principle, proceed via two mechanisms:  At tree level, {\em direct production} involves diagrams with two incoming partons, and two partons, a Higgs boson, and a photon in the final state (e.g.\ subprocesses of the type $qq'\to qq'H\ga$). However, rather than being directly produced, a photon in the final state can also stem from the {\em fragmentation} of a parton. This production mode involves diagrams with two partons in the initial state that produce a Higgs boson and three partons with one of these fragmenting into a photon, involving contributions like $qq'\to qq'Hg$ followed by the fragmentation of  a quark or gluon into a photon. To account for the non-perturbative nature of the fragmentation process in a fixed-order perturbative calculation, parton-to-photon fragmentation functions have to be employed. 
In multi-purpose Monte-Carlo generators, the fragmentation process can instead be simulated by non-perturbative hadronization models.  
In general, the description of fragmentation contributions poses various challenges: It involves non-perturbative fragmentation functions or hadronization models that are only poorly constrained. Moreover, within a Monte-Carlo generator including parton-shower effects it requires 
a careful combination of competing contributions associated with different Born structures matched to QED and QCD showers, respectively. In the \PBOX{} framework this issue has been addressed for prompt photon~\cite{Jezo:2016ypn} and $W\gamma$ production~\cite{Barze:2014zba}. Detailed studies on an appropriate treatment of photons have also been performed for the \HERWIG{}~\cite{DErrico:2011cgc} and \SHERPA{}~\cite{Siegert:2016bre} Monte-Carlo generators. 

To avoid the obstacles related to the computation and implementation of fragmentation contributions,   
{\em photon isolation} strategies have been devised by both experimentalists and theorists to suppress fragmentation contributions in such a way that the direct contributions provide a suitable description of a specific process involving a hard, identified photon.   
While differing in detail, all of these criteria rely on the idea of restricting the maximally allowed amount of transverse energy in a cone around the photon. 
The total transverse hadronic energy in a cone of size $r$ around a given  photon is defined as 
\begin{equation}
\label{eq:et-had}
E_T(r)=\sum_i E_{T,i} \, \Theta(r-\Delta R_{\gamma i}),
\end{equation}
with $E_{T,i}$ denoting the transverse energy of parton $i$ and $\Delta R_{\gamma i}$ the distance of parton~$i$ from the photon in the pseudorapidity-azimuthal angle plane. The summation in Eq.~\eqref{eq:et-had} extends over all partons around the photon.

The so-called \emph{fixed-cone} isolation criterion requires the total transverse hadronic energy within a cone of fixed size $r_0$ around a photon of transverse momentum $p_{T,\gamma}$ to be below some maximum value  $E_T^{max}$ \cite{Amoroso:2020lgh}, 
\begin{equation}
\label{eq:fcone-isolation}
E_T(r_0) \leq E_T^{max}
= \epsilon \, p_{T,\gamma} + E_T^{th},
\end{equation}
where the threshold transverse energy $E_T^{th}$, $\epsilon$ and the cone-size $r_0$ are free parameters. 
The \emph{smooth-cone} isolation criterion~\cite{Frixione:1998jh} extends the condition of Eq.~\eqref{eq:fcone-isolation} by introducing a profile function $\chi (r;r_0)$ and requiring
\begin{equation}
\label{eq:frixione-isolation}
E_T(r) \leq E_T^{max} \chi (r;r_0), \quad \forall r \leq r_0.
\end{equation}
While there is some freedom in the choice of $\chi (r;r_0)$, we follow the standard definition of
\begin{equation}\label{eq:profile-func}
\chi (r;r_0)=\left(\dfrac{1-\mathrm{cos}(r)}{1-\mathrm{cos}(r_0)} \right)^n,
\end{equation}
with a
parameter $n$  to be chosen by the user.
The \emph{hybrid-cone} isolation criterion \cite{Chen:2019zmr,Siegert:2016bre} combines the previous two prescriptions.
It is based on two cone criteria which the photons have to pass --first one inner smooth cone with radius $r_d$ and afterwards a larger fixed cone around the inner cone with $r_{0}> r_d$. Summarizing, we require 
\begin{align}\label{eq:hybrid-isolation}
E_T(r) &\leq E_T^{max} \chi (r;r_d), \quad \forall r \leq r_d, \nonumber\\
E_T(r_0) &\leq E_T^{max}
= \epsilon \, p_{T,\gamma} + E_T^{th}.
\end{align}

The validity of these approaches to isolate the direct photon production component has been explored in the literature (see, e.g., \cite{Gehrmann:2013aga}) and found to be satisfactory for selected applications.  For instance, a recent study~\cite{ATLAS:2023yrt} explored in detail the impact of different photon isolation criteria on the agreement between data and theoretical predictions for inclusive photon production. 
In the following we thus constrain ourselves to the calculation and simulation of the direct $\hajj$ production process. We will explore the uncertainty associated with the photon isolation procedure by comparing predictions with different isolation prescriptions. 

\begin{figure}[t] 
\begin{center}
\includegraphics[width=0.8\textwidth,clip]{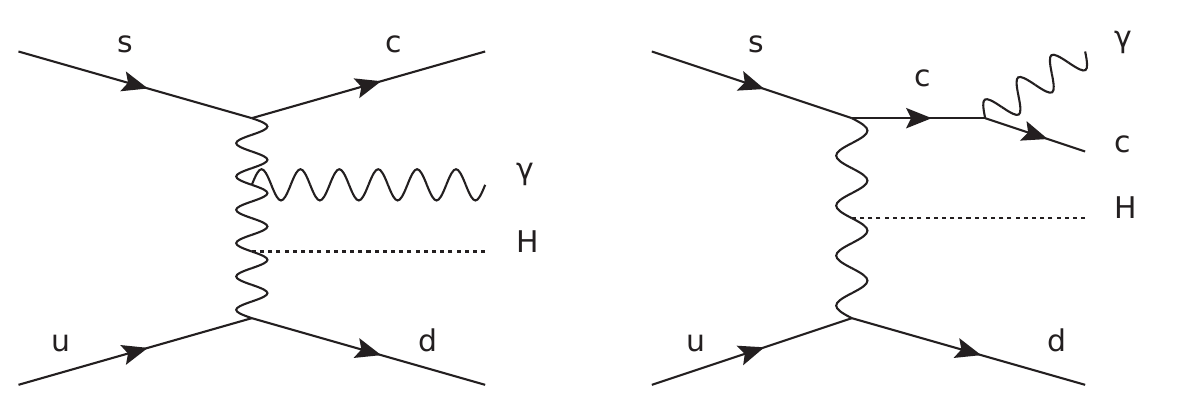}
\caption{Representative Feynman diagrams contributing to the VBF-induced partonic subprocess $us\to dcH\gamma$ at LO. 
}
\label{fig:born-diagrams}
\end{center}
\end{figure}

The purely EW VBF-induced $\hajj$ process at leading order (LO) involves the emission of a Higgs boson from a $Z$ or $W$~boson exchanged between the scattering (anti-)quarks. The accompanying hard photon can be emitted either from a quark line or, in charged-current subprocesses, from the exchanged $W$~boson, c.f.\ Fig.~\ref{fig:born-diagrams} for some representative diagrams. 
%
%
%
Depending on the partonic subprocess, either $t$- or $u$-channel topologies occur. In the case of same-flavor channels (such as $u u \to u u H \gamma$) both types of contributions arise. In such cases, we do take both of them individually into account, but neglect their interferences. 
At the same order in the EW coupling a gauge-invariant set of $s$-channel diagrams contributes to the $\hajj$ final state. Within the so-called {\em VBF approximation} this class of diagrams is neglected as well as interferences between the $t$- and $u$~channel topologies. The neglected contributions have been shown to be small for a large class of VBF and vector boson scattering (VBS) processes in the phase-space regions where they are experimentally explored, involving tagging jets of large invariant mass and rapidity separation \cite{Oleari:2003tc,Ciccolini:2007ec}. We will thus retain the VBF approximation for this work and refer to the $\hajj$ process within this approximation as {\em VBF $\hajj$ production}.  

\begin{figure}[t] 
\begin{center}
\includegraphics[width=0.8\textwidth,clip]{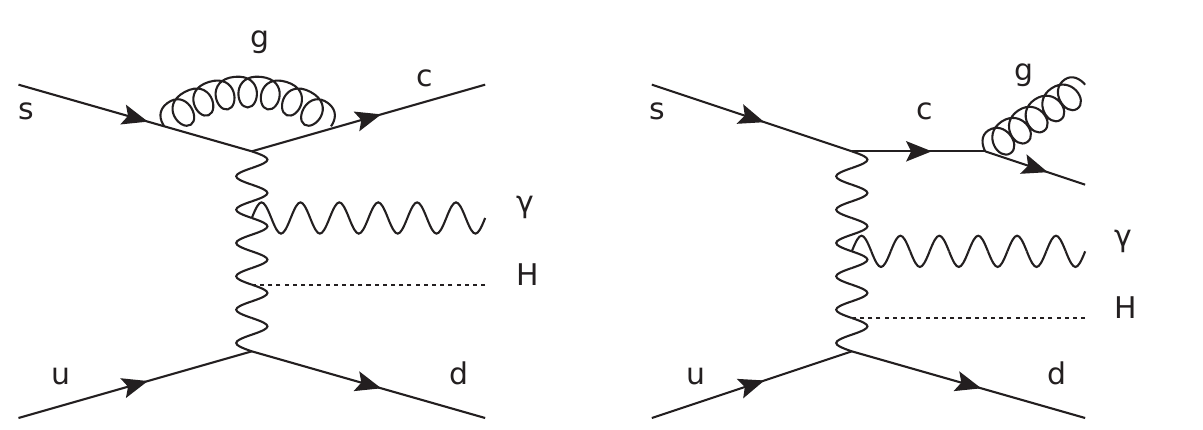}
\caption{Representative Feynman diagrams contributing to the VBF-induced partonic subprocess $us\to dcH\gamma$ at NLO-QCD level. 
}
\label{fig:nlo-diagrams}
\end{center}
\end{figure}

At the next-to-leading order (NLO) in QCD two classes of corrections arise: Real-emission contributions involving subprocesses with four external (anti-)quarks and an additional gluon in the final state, such as $qq'\to qq'gH\ga$, and crossing-related channels with a gluon in the initial state, for instance $gq'\to q\bar q q'H\ga$; virtual contributions involve one-loop corrections to either the upper or the lower quark line, interfered with the respective Born amplitude. Loop diagrams where a gluon is exchanged between the upper and the lower quark line do not occur within the VBF approximation.  Representative real-emission and virtual diagrams are depicted in Fig.~\ref{fig:nlo-diagrams}. 
%
%
%

For the implementation of the VBF $\hajj$ process within the \PBOXRES{}~\cite{Alioli:2010xd,Jezo:2015aia} the relevant partonic matrix elements have to be provided in a suitable format. We extracted the Born matrix elements,  real-emission and virtual QCD corrections from the parton-level implementation of Ref.~\cite{Arnold:2010dx}.  
The color- and spin-correlated Born amplitudes needed for the computation of infrared subtraction terms can be obtained, with suitable modifications to account for the VBF approximation,  from a build tool based on {\tt MadGraph4}~\cite{Alwall:2007st} that is available within the public version of the \PBOX{}.  
To be as generic as possible, we do not focus on a particular decay mode of the Higgs boson. When our \PBOX{} implementation is combined with a multi-purpose generator such as \PYTHIA{}, any tree-level decay can be simulated in a straightforward manner. 

For the phase-space integration we developed a customized parameterization inspired by the \PBOX{} implementation of the related VBF-induced $\hjjj$ process~\cite{Jager:2014vna}. 
To improve the numerical stability of the phase-space integration we apply a  minimal set of cuts at the generation level: 
The transverse momenta of the two final-state quarks and the photon of the underlying Born configuration $\Phi$, i.e.\ $p_{T,f_i}$ with $i=1,2$  and $p_{T,\gamma}$, are required to be in the range 
\begin{equation}
p_{T,f_i} > 10~\mathrm{GeV}, \quad 
p_{T,\gamma} > 10~\mathrm{GeV}.
\end{equation}
Additionally, the invariant mass and separation in pseudorapidity of the final-state quark-pair has to satisfy the constraints 
\begin{equation}
m_{f_1 f_2} > 30~\mathrm{GeV}, \quad 
\vert \eta_{f_1}-\eta_{f_2} \vert > 1.
\end{equation}
The final-state quarks are furthermore required to be well separated from the photon in the pseudorapidity-azimuthal angle plane, 
\begin{equation}
\Delta R_{f\gamma} > 0.1,
\end{equation} 
with scalar products fulfilling
\begin{equation}
2 p_{f_i} \cdot p_{\gamma} >  0.1~\mathrm{GeV}.
\end{equation}

In addition to these generation cuts we employ a Born-suppression factor of the form 
\begin{equation}
F(\Phi)= \left( \dfrac{p_{T,f_1}^2}{p_{T,f_1}^2+\Lambda_{\Phi}^2}\right)^k \left( \dfrac{p_{T,f_2}^2}{p_{T,f_2}^2+\Lambda_{\Phi}^2}\right)^k \left( \dfrac{p_{T,\gamma}^2}{p_{T,\gamma}^2+\Lambda_{\Phi}^2}\right)^k,
\end{equation}
where $k=2$ and $\Lambda_{\Phi}=10~\mathrm{GeV}$ are technical parameters. We verified that numerical results are independent of the specific values of these parameters within statistical uncertainties. 

We checked that the Born and real-emission amplitudes within our \PBOXRES{} implementation agree with respective amplitudes generated with \MG{}~\cite{Stelzer:1994ta} for individual phase-space points up to ten digits. Cross sections within the VBF-specific cuts applied in our phenomenological analysis have been compared 
with those obtained with the  \VBFNLO{} parton-level Monte-Carlo generator~\cite{Arnold:2008rz,Baglio:2011juf} and with \MGNLO~\cite{Alwall:2014hca}.  
We found agreement within the numerical accuracy of the programs. 

\section{Phenomenological results}
\label{sec:pheno}
Let us now explore the impact of parton-shower effects and different photon isolation prescriptions on the VBF $\hajj$ process. 
\subsection{Setup}
We consider proton-proton collisions at the LHC with a center-of-mass energy of $\sqrt{s}=13$~TeV.
Throughout we use the $G_{\mu}$ input scheme for EW parameters with a Fermi constant of $G_\mu=1.16639 \times 10^{-5} \, \mathrm{GeV}$. The masses and widths of the EW bosons are set to~\cite{ParticleDataGroup:2024cfk}
\bea
m_Z&=& 91.1880 \, \gev, \quad
\Gamma_Z= 2.4955 \, \gev, 
\\
m_W&=& 80.3692 \, \gev,\quad 
\Gamma_W= 2.085 \, \gev. 
\\
m_H&=& 125.20 \, \gev,\quad 
\Gamma_H= 3.7 \, \mev. 
\eea 

We assume five active quark flavors and  a diagonal form of the Cabibbo-Kobayashi-Maskawa matrix. 
For the parton distribution functions (PDFs) we use the PDF4LHC21\_40 set as provided by version 6.5.3 of the \texttt{LHAPDF} library~\cite{Buckley:2014ana} together with the corresponding value of the strong coupling $\alpha_s$. 
Jets are reconstructed with the anti-$k_T$ algorithm~\cite{Cacciari:2008gp} as implemented within the {\tt FastJet} tool~\cite{Cacciari:2011ma} (version 3.4.3), with a radius-parameter of $R=0.4$.

For the renormalization scale, $\mur=\xir\mu_0$, and factorization scale, $ \muf=\xif\mu_0$, we use 
\begin{equation}
\mu_0^2=\dfrac{m_H}{2}\sqrt{\dfrac{m_H^2}{4}+p_{T,H}^2}.
\end{equation}
Below, the factors $\xir$ and $\xif$ are varied between $0.5$ and $2$ with their ratios   constrained to the range 0.5 to 2 (in the following referred to as 7-point variation) to estimate the associated theoretical uncertainties. 

For the simulation of parton-shower (PS) effects we use the QCD shower of  \texttt{PYTHIA} version~8.312~\cite{bierlich2022comprehensiveguidephysicsusage} with the \texttt{Monash 2013} tune~\cite{Skands_2014}. Following recommendations for the proper simulation of VBF processes in the literature~\cite{Hoche:2021mkv,Barone:2025jey}, we avoid using the default global shower, but rather employ the dipole-local shower within \PYTHIA{}. We also show results obtained with the antenna shower \VINCIA~ \cite{Fischer:2016vfv} included in \texttt{PYTHIA}, and by \HERWIGS{} version 7.3.0 \cite{Bewick:2023tfi}. Unless specifically stated otherwise, multi-parton interactions (MPI), underlying event (UE), and hadronization effects are deactivated. QED shower effects are specifically not taken into account, as in our Monte-Carlo generator they would require a different treatment of the hard photon that is part of the $\hajj$ signal process. 

For our phenomenological studies we employ a set of selection cuts to ensure the validity of the VBF approximation. Our analysis setup is inspired by typical selection criteria used in experimental analyses for this process~\cite{ATLAS:2020cvh}.
We require the presence of at least two jets with transverse momenta and rapidities in the range 
\beq
\label{eq:cuts-jets}
p_{T,j} > 40 \, \mathrm{GeV}, \quad \vert \eta_j \vert < 4.5\,.
\eeq
The two hardest jets fulfilling these criteria are referred to as {\em tagging jets}. These additionally have to exhibit a large invariant mass and rapidity separation,
\beq
\label{eq:cuts-tag}
m_{jj}^{tag} > 800~\gev, \quad 
\Delta \eta_{jj}^{tag} = \vert \eta_{j_1}^{tag} - \eta_{j_2}^{tag} \vert > 2.5\,. \eeq
When considering observables related to subleading jets we only consider candidates with a  minimum transverse momentum and maximum pseudorapidity of 
\beq
\label{eq:cuts-jet3}
p_{T,j}^{sub} > 10 \, \mathrm{GeV}\,, \qquad \vert \eta_j^{sub} \vert < 4.5\,. 
\eeq
In cases where we consider specifically the transverse momentum of non-tagging jets the $p_{T,j}^{sub}$ requirement is lifted. 

Additionally we apply a cut on the transverse momentum of the photon
\beq
\label{eq:cuts-photon}
p_{T,\gamma} > 30 \, \mathrm{GeV},\,\quad  \vert \eta_{\gamma}\vert < 2.4 \,.
\eeq

For the photon-isolation schemes described in Eqs.~(\ref{eq:fcone-isolation}-\ref{eq:hybrid-isolation}) we use the following parameters,
\beq\label{eq:iso-parameters}
E_T^{th} = 4.8\, \mathrm{GeV}\,, \quad \epsilon = 1\,, \quad n = 1\,, \quad r_0 = 0.4\,, \quad r_d = 0.1 \,,
\eeq
unless specified otherwise. We use the smooth-cone isolation algorithm as the standard option in our NLO matched to parton shower (NLO+PS) discussion. 


\subsection{Assessment of parton shower dependencies} 

In this section we present results at NLO+PS accuracy with \PYTHIAE{} and \HERWIGS{} for a number of kinematic observables.

\begin{figure}[pt!]
\centering
\subfloat[][]{
\includegraphics[width=0.5\textwidth]{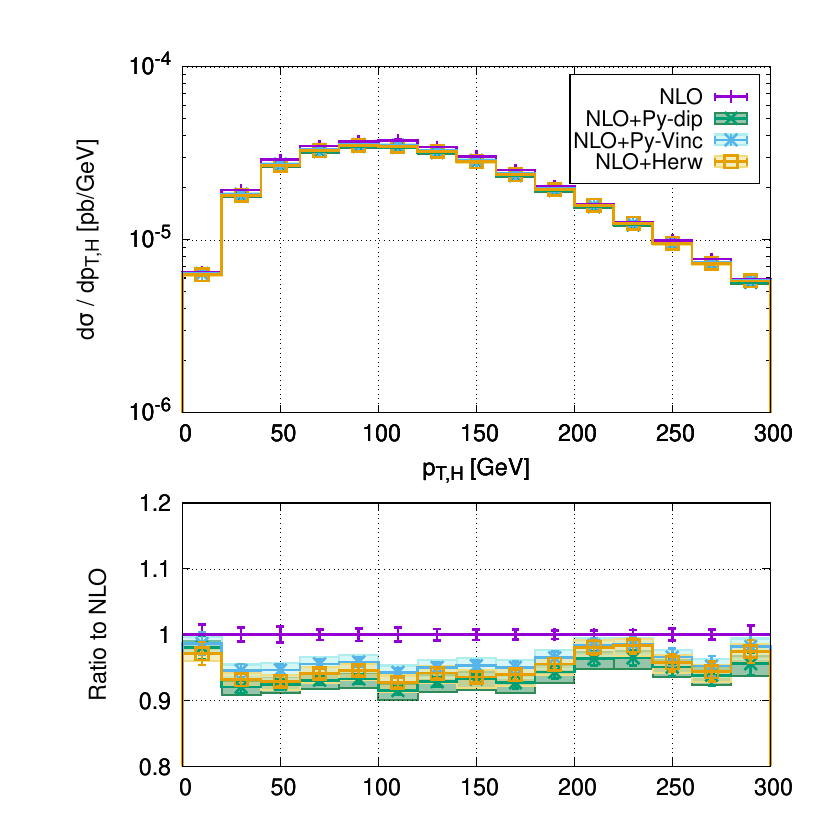}}
\subfloat[][]{
\includegraphics[width=0.5\textwidth]{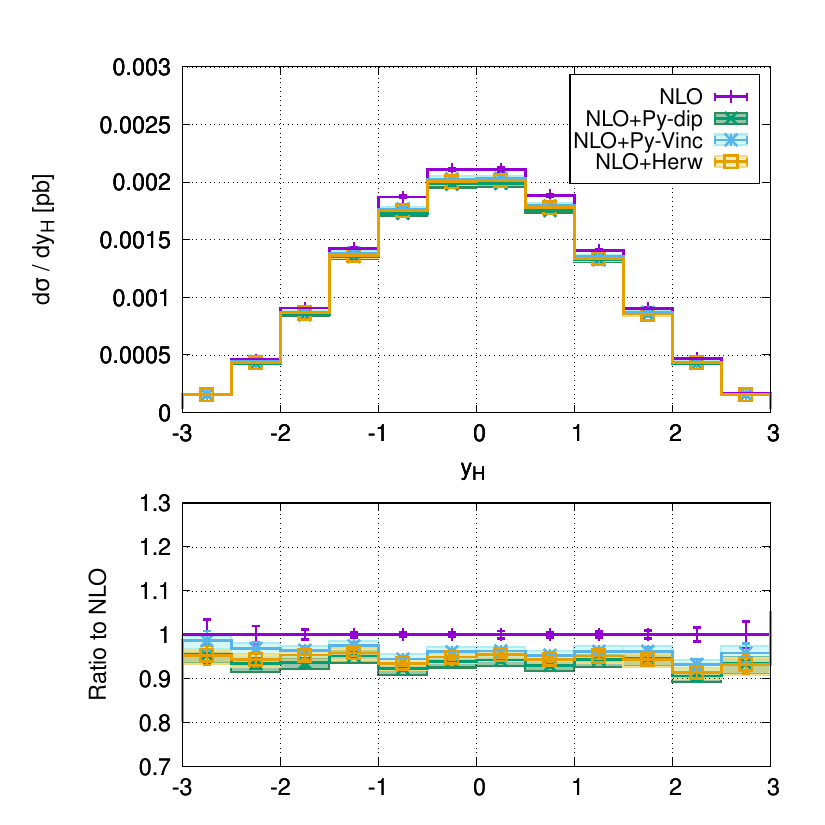}}\\
\subfloat[][]{
\includegraphics[width=0.5\textwidth]{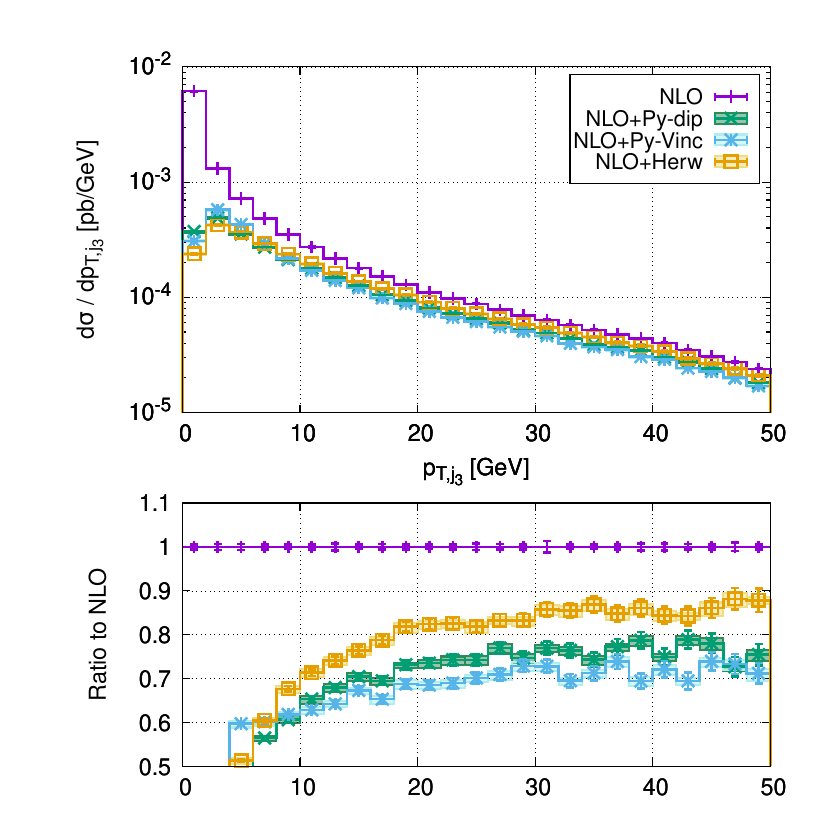}}
\subfloat[][]{
\includegraphics[width=0.5\textwidth]{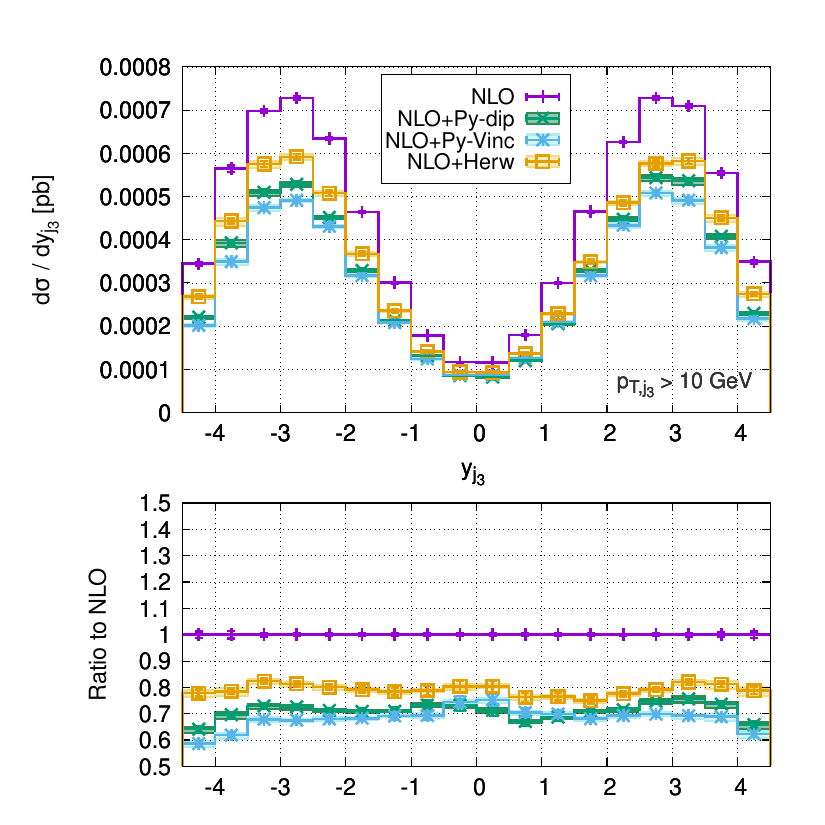}}
\caption{\label{fig:her-vinc} Transverse momentum and rapidity distributions of the Higgs boson (top) and the third hardest jet (bottom) at  NLO (purple), NLO+\PYTHIA{} with the dipole-recoil scheme (green), NLO+\VINCIA{} (blue) and NLO+\HERWIG{} (yellow). In the panels below ratios to NLO are shown. Error bars indicate statistical uncertainties and we show uncertainty bands for 7-point scale variation.} 
\end{figure}

In Fig.~\ref{fig:her-vinc} the transverse momentum and rapidity of the Higgs boson and of the third hardest jet are shown at NLO, at NLO+\PYTHIA{} with the dipole-recoil scheme, NLO+\texttt{VINCIA} and NLO+\HERWIG{}. 
%
%
%
%
We observe good agreement between the different shower algorithms for the observables regarding the Higgs boson. In general, the NLO+PS results tend to lie slightly below the NLO results, with differences of at most 9\%. For third-jet observables we see larger differences. Since the third jet only enters the NLO calculation via real-emission contributions it is essentially only treated at tree-level within our NLO calculation, resulting in a lower perturbative accuracy for observables related to the third hardest jet than typical for observables accounted for with genuine NLO accuracy. This results in larger differences between NLO and NLO+PS predictions. Furthermore, the NLO prediction exhibits a divergent behavior for $p_{T,j_3} \rightarrow 0$, which is dampened by Sudakov suppression in the hardest-emission generation of the \PBOX.

\begin{figure}[pt!]
\centering
\subfloat[][]{
\includegraphics[width=0.5\textwidth]{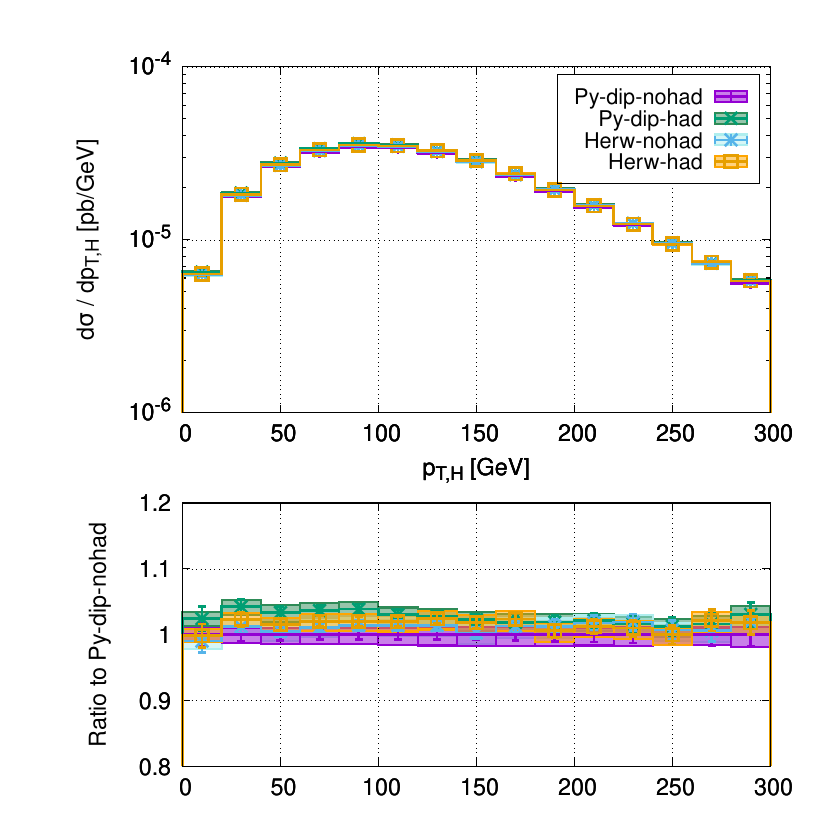}}
\subfloat[][]{
\includegraphics[width=0.5\textwidth]{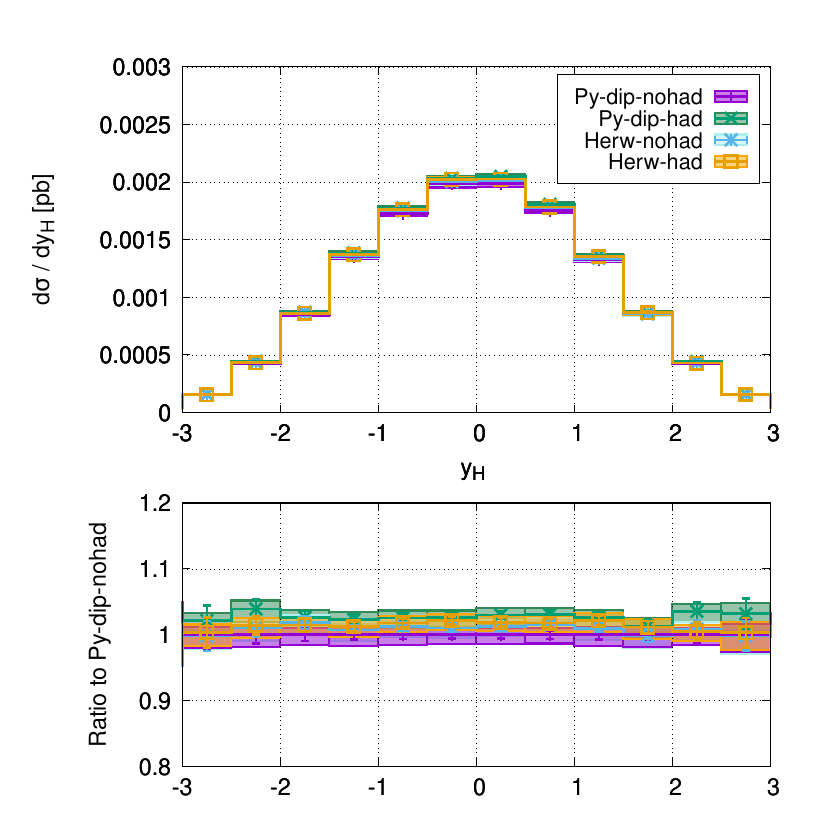}}\\
\subfloat[][]{
\includegraphics[width=0.5\textwidth]{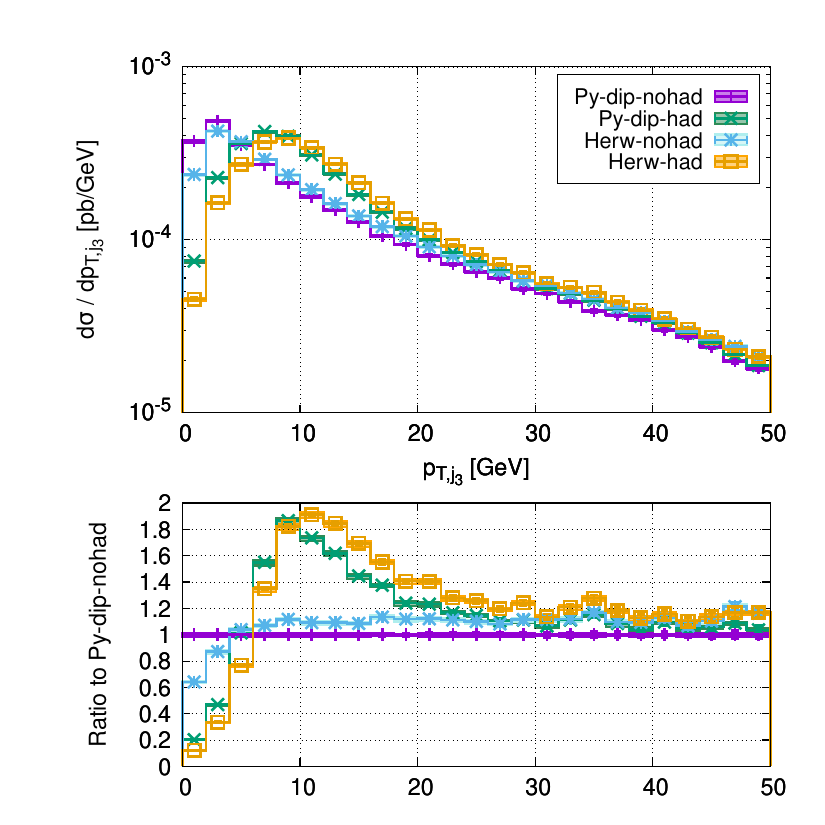}}
\subfloat[][]{
\includegraphics[width=0.5\textwidth]{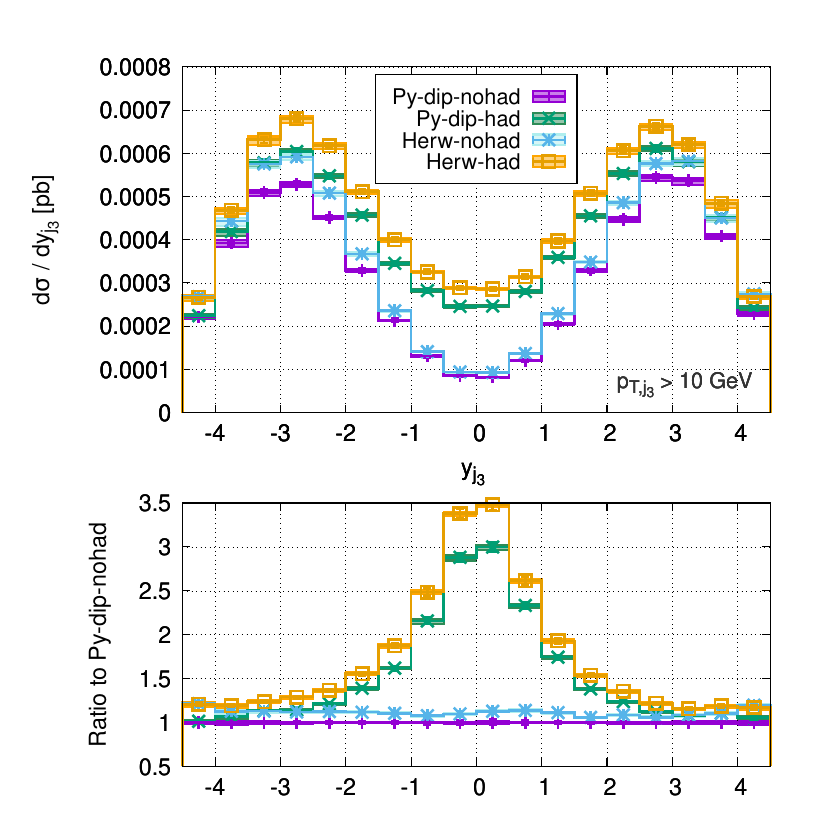}}
\caption{\label{fig:had} Transverse momentum and rapidity distributions of the Higgs boson (top) and the third hardest jet (bottom) at NLO+\PYTHIA{} with the dipole-recoil scheme without (purple) and with (green) non-perturbative effects, and NLO+\HERWIG{} without (blue) and with non-perturbative effects (yellow). Error bars indicate statistical uncertainties and we show uncertainty bands for 7-point scale variation.}
\end{figure}

In Fig.~\ref{fig:had}, 
%
%
%
%
we show the impact of hadronization effects, MPI and underlying event (in the following collectively referred to as {\em non-perturbative effects}) on the same observables. 
We see that for Higgs boson observables these 
effects do not have a large impact on the distributions. Differences amount to a few percent at most and are basically contained within the statistical uncertainties. Third jet observables are much more sensitive to non-perturbative effects, and we observe large differences with respect to the purely perturbative predictions  for both \PYTHIA{} and \HERWIG{} in the regime of low $p_{T,j_3}$ and $y_{j_3}$ values. Especially the region around zero rapidity is filled by hadronization, MPI, and UE. Similar effects have been observed before for other VBF processes \cite{Jager:2018cyo,Bittrich:2021ztq,Jager:2025isz}. This behavior can be consistently observed in both generators \PYTHIA{} and \HERWIG{}. We note that the changes in shape at low transverse momenta and rapidities mostly stem from UE and MPI effects, 
while hadronization effects only kick in for transverse momenta larger than 5~GeV.

%
\subsection{Impact of the photon-isolation algorithm}
\begin{figure}[pt!]
\centering
\subfloat[][]{
\includegraphics[width=0.5\textwidth]{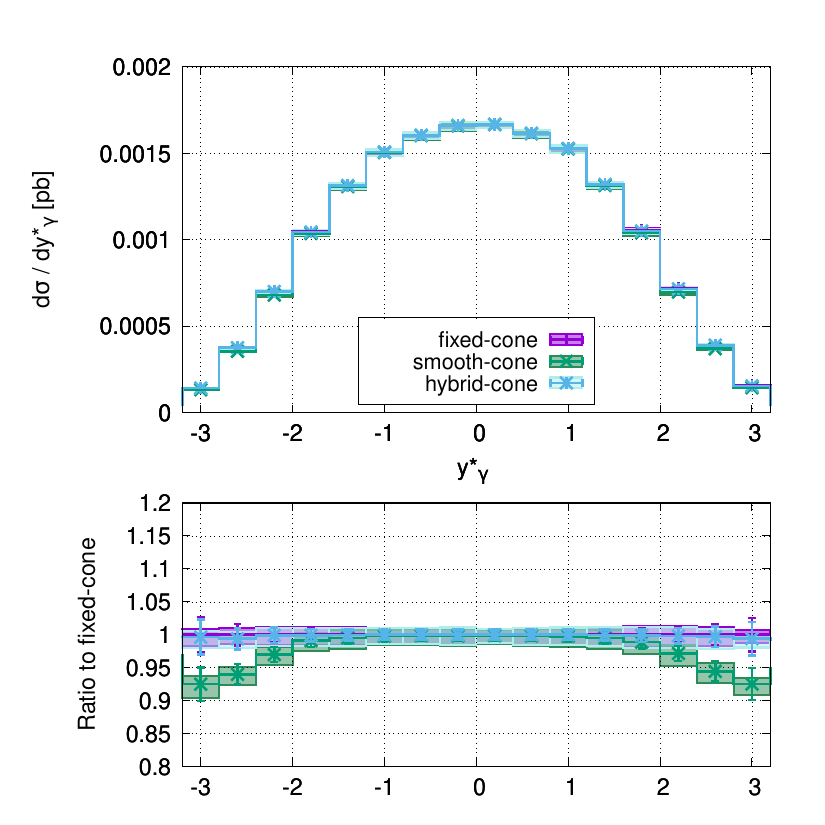}}
\subfloat[][]{
\includegraphics[width=0.5\textwidth]{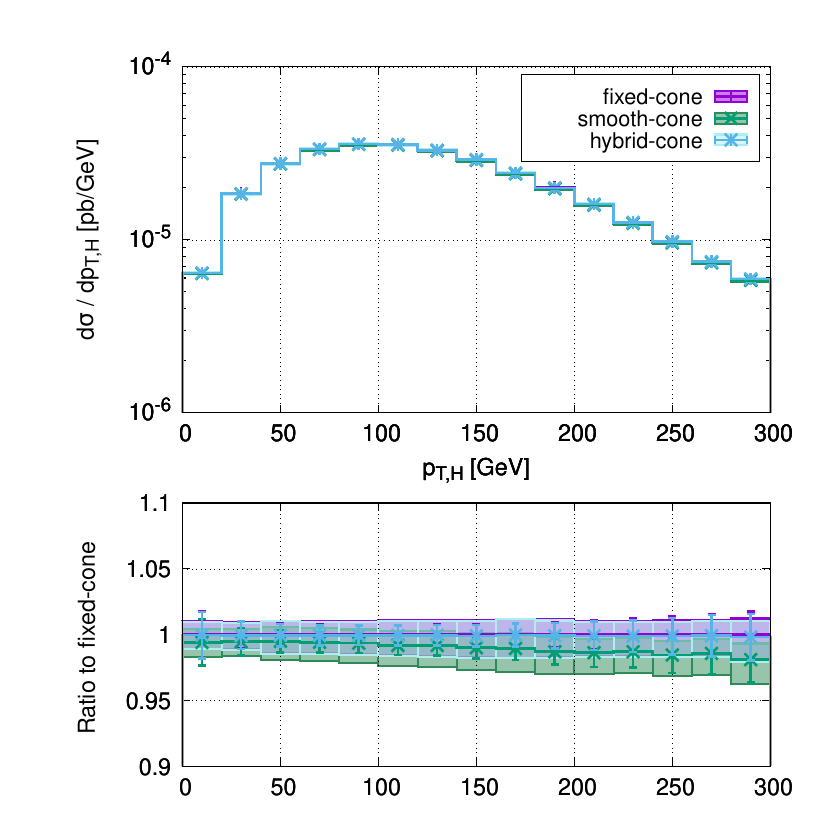}}
\caption{\label{fig:photo-iso-comp} Distributions of $y^{*}_{\gamma}$ (left) and $p_{T,H}$ (right) for different photon-isolation algorithms: fixed-cone algorithm (purple), smooth-cone algorithm (green), and hybrid-cone algorithm (blue). The lower panels show ratios to results with the fixed-cone algorithm. Error bars indicate statistical uncertainties, and we show uncertainty bands for 7-point scale variation.}
\end{figure}

In this subsection we discuss the impact of the different photon-isolation algorithms on the $\hajj$ signal signature, comparing the standard, the smooth cone, and the hybrid-cone isolation prescriptions. 
To that end, we specifically consider an observable constructed from the position of the photon relative to the center of the two tagging-jet system, 
\begin{equation}
\label{eq:ystargamma}
y^{*}_{\gamma}=y_{\gamma} - \dfrac{y_{j_1}+y_{j_2}}{2}\,.
\end{equation}
In Fig.~\ref{fig:photo-iso-comp}
%
%
%
%
we compare results for $y^{*}_{\gamma}$ using various isolation prescriptions. 
We find that the hybrid-cone algorithm yields results very similar to the fixed-cone algorithm with differences being fully contained in the respective uncertainties. The smooth cone algorithm exhibits small deviations for larger absolute values of $y^{*}_{\gamma}$, with results lying a few percent below the ones for the fixed-cone algorithm in these regions. Since $y^{*}_{\gamma}$ is a measure for how far away the photon lies from the center of the tagging-jet system, the choice of photon-isolation algorithm is expected to have the largest impact in the large $y^{*}_{\gamma}$ regions. In VBF processes,  hadronic activity typically is located close to the tagging jets rather than in between them, a feature that is also exploited in central-jet veto techniques~\cite{Rainwater:1999sd}.
For non-photon related observables, such as the transverse momentum of the Higgs boson also displayed in Fig.~\ref{fig:photo-iso-comp},  we observe only minimal changes. For those types of distributions, effects of the different isolation algorithms are fully contained within the respective statistical uncertainties. 
Our findings on the rather small sensitivity of predictions on the photon isolation prescription  of even the most scheme-dependent observables in the VBF-initiated $\hajj$ process indicate that details of how the direct photon component is extracted from the full photon-production process in this channel have little phenomenological impact. 

\begin{figure}[pt!]
\centering
\subfloat[][]{
\includegraphics[width=0.5\textwidth]{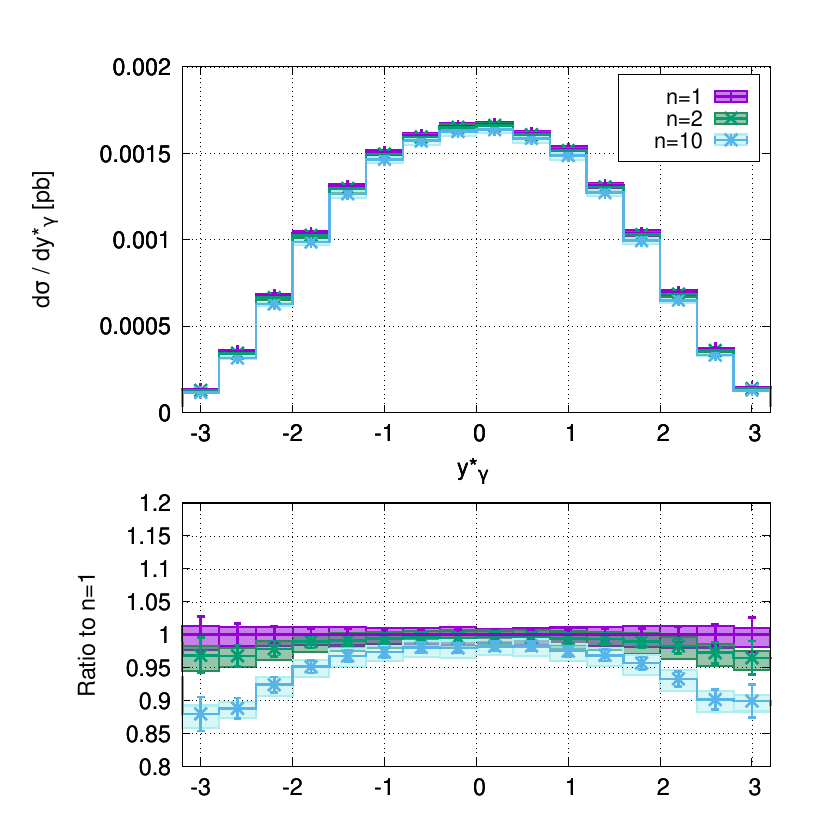}}
\subfloat[][]{
\includegraphics[width=0.5\textwidth]{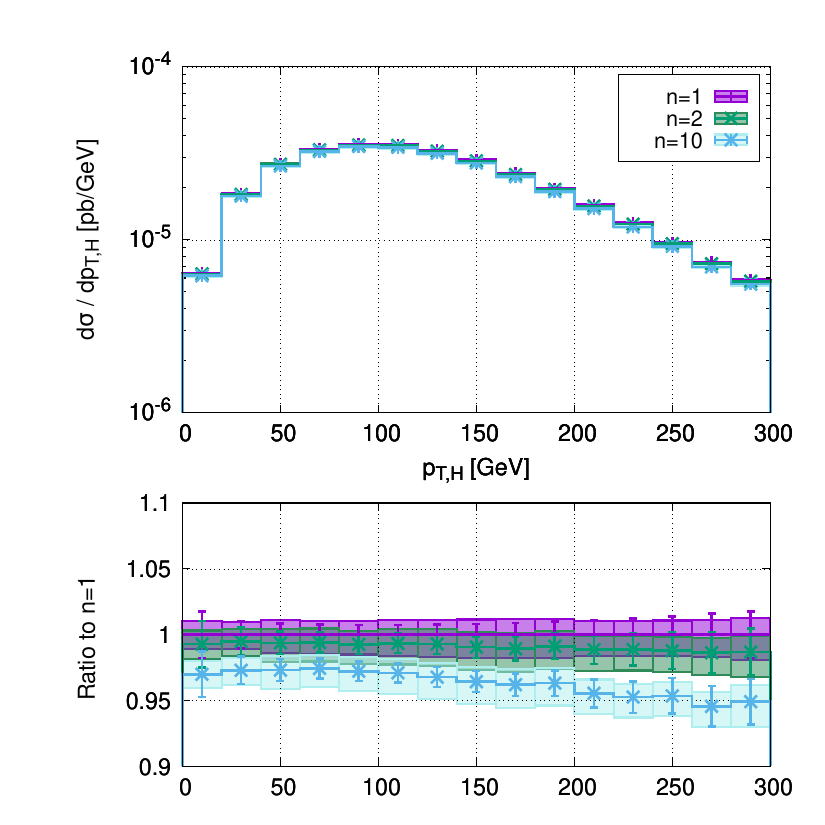}}
\caption{\label{fig:photo-iso-n} Distributions of $y^{*}_{\gamma}$ (left) and $p_{T,H}$ (right) with the smooth photon-isolation algorithm for $n=1$ (purple), $n=2$ (green) and $n=10$ (blue). The lower panels show ratios to $n=1$. Error bars indicate statistical uncertainties, and we show uncertainty bands for 7-point scale variation.}
\end{figure}

To further explore the dependence of predictions on the photon isolation prescription, 
in Fig.~\ref{fig:photo-iso-n} 
%
%
%
%
we show results for the smooth-cone isolation scheme with different values of the parameter $n$ of Eq.~(\ref{eq:profile-func}).  We refrain from showing similar results  for the hybrid cone algorithm, since there the impact of the parameter $n$ is negligible due to the fixed-cone within the hybrid-scheme dominating the photon selection.

We observe an amplification of the differences of the smooth cone algorithm compared to the fixed-cone isolation in the large $y^{*}_{\gamma}$ region with increasing values of $n$. We also see a dependence on the parameter for other observables not as sensitive to the photon-isolation algorithm. For $n=10$ we observe a shift downwards of $p_{T,H}$ of up to 5\%, due to a lower amount of events passing the photon-selection procedure. 
We want to stress that a rather extreme value of $n=10$ is chosen here only to highlight the impact of the parameter on observables explicitly sensitive to the isolation algorithm. For more reasonable values of $n$ and considering other observables, the overall impact of $n$ on the results is  small.

\begin{figure}[pt!]
\centering
\subfloat[][]{
\includegraphics[width=0.5\textwidth]{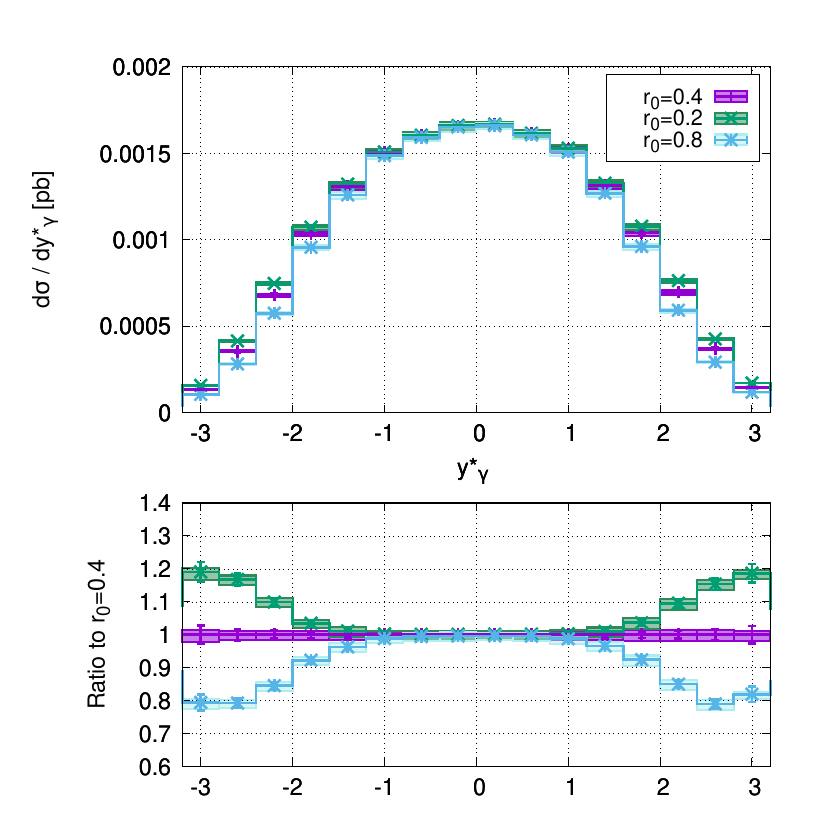}}
\subfloat[][]{
\includegraphics[width=0.5\textwidth]{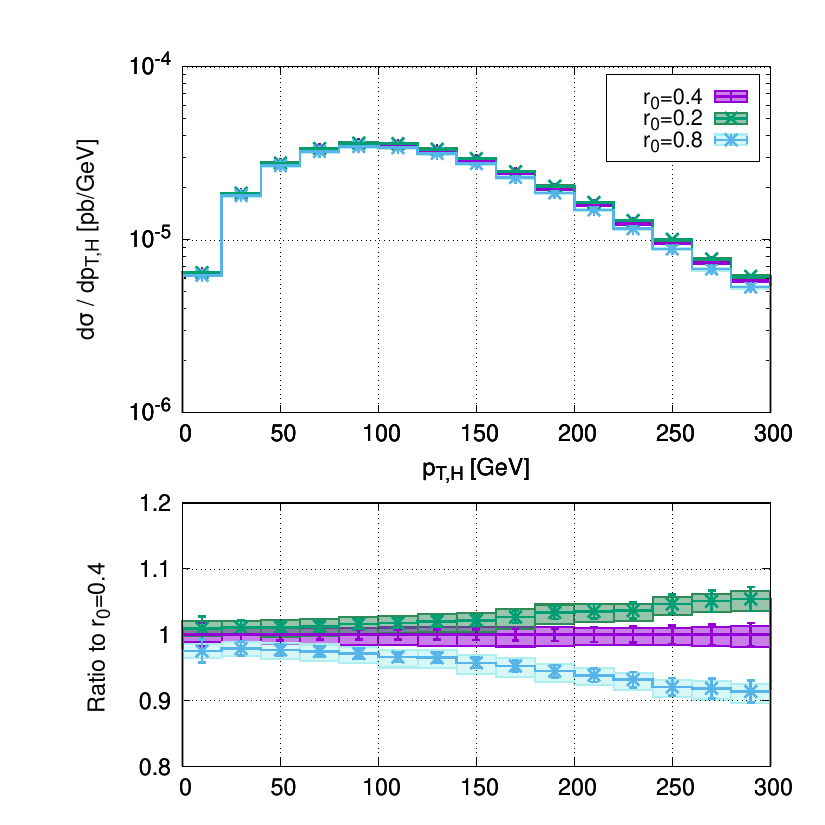}}
\caption{\label{fig:photo-iso-r} Distributions of $y^{*}_{\gamma}$ (left) and $p_{T,H}$ (right) with the smooth photon-isolation algorithm for $r=0.4$ (purple), $r=0.2$ (green) and $r=0.8$ (blue). The lower panels show ratios to $r=0.4$. Error bars indicate statistical uncertainties, and we show uncertainty bands for 7-point scale variation.}
\end{figure}

Finally,  in Fig.~\ref{fig:photo-iso-r}  we consider the dependence on the size of the cone $r_0$ within the smooth-cone isolation scheme. 
%
%
%
%
Similar to the discussion of $n$, we see larger differences of up to 20\% in the large $y^{*}_{\gamma}$ region. A smaller cone-radius $r_0$ leads to an increase, a larger one to a decrease in these regions. A qualitatively similar  behavior can be observed for the Higgs transverse momentum, but with smaller impact. 
This trend was to be expected since a larger isolation-cone radius results in more events with hadronic activity in this cone being discarded, reducing the cross section in these regions. In contrast, a smaller isolation-cone radius leads to more events passing the isolation criteria, and therefore a larger cross section.

\section{Summary and conclusions}
\label{sec:conclusions}
In this work we presented an implementation of the VBF-induced $\hajj$ process in the framework of the \PBOXRES. The Monte-Carlo program we developed is publicly available from 
\url{https://gitlab.com/POWHEG-BOX/RES/User-Processes/VBF_Hgamma}. 

We studied the impact of different parton showers finding good agreement for Higgs boson observables, while larger differences occur for distributions of subleading jets.  
We observed that hadronization, MPI, and UE significantly modify the transverse momentum and rapidity distributions of the hardest subleading jet, increasing the number of events at central rapidities. Such a behavior has also been observed for other VBF and VBS processes ~\cite{Jager:2018cyo,Bittrich:2021ztq,Jager:2025isz}. 

Various strategies for isolating the direct photon production component were explored, in particular a fixed-cone, a smooth-cone, and a hybrid cone approach. For typical experimental setups the different algorithms resulted in very similar predictions, with differences of at most 5\% in the region of large $y^{*}_{\gamma}$ region where the isolation algorithm has the largest impact.  

For setups with unphysical values of the cone parameters we instead observed larger differences also for observables not directly related to the photon, due to a different amount of events passing the photon-isolation cuts. 
We therefore advise practitioners to exercise care when choosing parameter values for the photon isolation prescription.

%
\section*{Acknowledgements}
We are grateful to Ignacio Borsa for helpful discussions and to Alexander Puck Neuwirth and Carlo Oleari for assistance with the \PBOX{} repository. 
This work has been funded by the Deutsche Forschungsgemeinschaft (DFG, German Research Foundation) -- Project No.~563181713. 

%

\bibliographystyle{JHEP}
\bibliography{vbfah}

\end{document}